\documentclass[aps,twocolumn,prl,superscriptaddress,showpacs,showkeys]{revtex4-1}
\usepackage{graphicx}
\begin{document}

\title{Transforming a Surface State of Topological Insulator by a Bi Capping Layer}

\author{Han Woong Yeom}
\email{yeom@postech.ac.kr}
\affiliation{Center for Artificial Low Dimensional Electronic Systems, Institute for Basic Science (IBS), 77 Cheongam-Ro, Pohang 790-784, Korea}
\affiliation{Department of Physics, Pohang University of Science and Technology, 77 Cheongam-Ro, Pohang 790-784, Korea}

\author{Sung Hwan Kim}
\affiliation{Center for Artificial Low Dimensional Electronic Systems, Institute for Basic Science (IBS), 77 Cheongam-Ro, Pohang 790-784, Korea}
\affiliation{Department of Physics, Pohang University of Science and Technology, 77 Cheongam-Ro, Pohang 790-784, Korea}

\author{Woo Jong Shin}
\affiliation{Center for Artificial Low Dimensional Electronic Systems, Institute for Basic Science (IBS), 77 Cheongam-Ro, Pohang 790-784, Korea}
\affiliation{Department of Physics, Pohang University of Science and Technology, 77 Cheongam-Ro, Pohang 790-784, Korea}

\author{Kyung-Hwan Jin}
\affiliation{Department of Physics, Pohang University of Science and Technology, 77 Cheongam-Ro, Pohang 790-784, Korea}

\author{Joonbum Park}
\affiliation{Department of Physics, Pohang University of Science and Technology, 77 Cheongam-Ro, Pohang 790-784, Korea}

\author{Tae-Hwan Kim}
\affiliation{Center for Artificial Low Dimensional Electronic Systems, Institute for Basic Science (IBS), 77 Cheongam-Ro, Pohang 790-784, Korea}
\affiliation{Department of Physics, Pohang University of Science and Technology, 77 Cheongam-Ro, Pohang 790-784, Korea}

\author{Jun Sung Kim}
\affiliation{Department of Physics, Pohang University of Science and Technology, 77 Cheongam-Ro, Pohang 790-784, Korea}

\author{Hirotaka Ishikawa}
\affiliation{Department of Nanomaterials Science, Chiba University, Chiba 263-8522, Japan}

\author{Kazuyuki Sakamoto}
\affiliation{Department of Nanomaterials Science, Chiba University, Chiba 263-8522, Japan}

\author{Seung-Hoon Jhi}
\affiliation{Department of Physics, Pohang University of Science and Technology, 77 Cheongam-Ro, Pohang 790-784, Korea}

\date{June 15, 2014}

\begin{abstract}
We introduce a dinstint approach to engineer a topologically protected surface state of a topological insulator. By covering the surface of a topological insulator, Bi$_2$Te$_2$Se, with a Bi monolayer film, the original surface state is completely removed and three new spin helical surface states, originating from the Bi film, emerge with different dispersion and spin polarization, through a strong electron hybridization. These new states play the role of topological surface states keeping the bulk topological nature intact. This mechanism provides a way to create various different types of topologically protected electron channels on top of a single topological insulator, possibly with tailored properties for various applications.
\end{abstract}

\maketitle

Topological insulators (TIs) are a new class of insulator materials with unusual surface (edge) metallic electron channels \cite{RMP.82.3045, RMP.83.1057}. These surface channels have massless Dirac electron character with helical spin polarization \cite{N.452.970, N.460.1101, S.325.178, N.466.343, PRL.104.016401, PRL.106.216803, NCom.3.636, S.329.659, S.339.1582, S.332.560}, which are robustly protected by the topological nature of the bulk \cite{NP.5.438, PRL.105.186801}. These unique characteristics make surface states of TI, topological surface states (TSS), ideal for scattering-free carriers of spintronic information and the fault-tolerant quantum computing.

The prompt application of these materials are, however, hampered by not only materials issues such as surface and bulk imperfections but also by the robustly protected nature of their surface states itself; intrinsically hard to manipulate and control \cite{PRL.105.186801}. So far, the most popular way of controlling a TSS is to dope them. For the cases of the most widely studied 3D TI of Bi chalcogenides, the non-magnetic atomic and molecular dopants \cite{N.460.1101, PRL.104.016401, S.325.178, S.332.560, NCom.3.636, NP.7.32} were shown to shift the TSS bands. On the other hand, the magnetic impurity atoms were debatably reported to open a small band gap at Dirac points of TSS by breaking the time reversal symmetry \cite{S.329.659, S.339.1582}. However, the topological property of the materials is, then destroyed and the magnetic impurities give rise to unwanted scatterings. Very recently, theoretical works suggested the changes in the effective mass of TSS upon terminating the surface with other atoms \cite{PRL.111.146803} and the changes in the vertical position and the Dirac point energy of TSS upon capping the surface with insulating ultrathin films \cite{acsnano.6.2345,tuning, band_engineering}. However, none of these proposals were realized.

In this Letter, we devise a distinct approach in engineering TSS. We show both theoretically and experimentally that a strongly interacting monolayer grown on top of a 3D TI, a simple Bi monolayer on Bi2Te2Se (called BTS hereafter), in particular, can host new spin-helical electronic states replacing the original TSS with the bulk TI property preserved. By changing the terminating monolayer, one can generate various different TSS with different dispersion and spin orientation on a single 3D TI. That is, a particular TSS can be replaced by or transformed into other helical Dirac electronic states satisfying the topological requirement of the system. This result opens a new avenue towards creating and tailoring topologically protected spin and electron channels.

We used cleaved single crystals of BTS, which were grown using the self flux methods \cite{PRB.82.241306,PRB.89.155436}. The BTS crystals were cleaved in ultra high vacuum, onto which Bi monolayers were deposited as reported before \cite{PRB.89.155436}. We performed scanning tunneling microscopy/spectroscopy
(STM/STS) measurements using a commercial low-temperature STM. The STM topography was measured in a constant-current mode, and STS spectra and maps were obtained by the lock-in technique, which minimize the effect of topographic corrugations with the current feedback turned off \cite{PRB.89.155436}. The spin- and angle-resolved photoemission spectroscopy (SARPES) measurements were performed with a high performance hemispherical electron analyzer (VG-SCIENTA R4000) and a Mott spin detector using Xe discharge light of 8.4 eV \cite{Ncom.4.2073}. The samples were kept at 78 K in both experiments.

The \textit{ab initio} calculations were carried out in the plane-wave basis within generalized gradient approximation for exchange-correlation functional \cite{PRB.54.11169, PRL.77.3865}. A cutoff energy of 400 eV was used for the plane-wave expansion. The Bi/Bi$_2$Te$_2$Se structure is simulated by the supercell with Bi one monolayer (in a bilayer structure) on one surface of a slab of six quintuple layers (QLs) Bi$_2$Te$_2$Se and a vacuum layer of 20\AA-thick between the cells \cite{PRB.89.155436}. During structural relaxation, the atoms of Bi monolayer and three Bi$_2$Te$_2$Se surface layers are allowed to relax until the forces are smaller than 0.01 eV/{\AA}.  The van-der Waals interaction is also considered \cite{PRL.102.073005}.

\begin{figure}[!pb]
\includegraphics{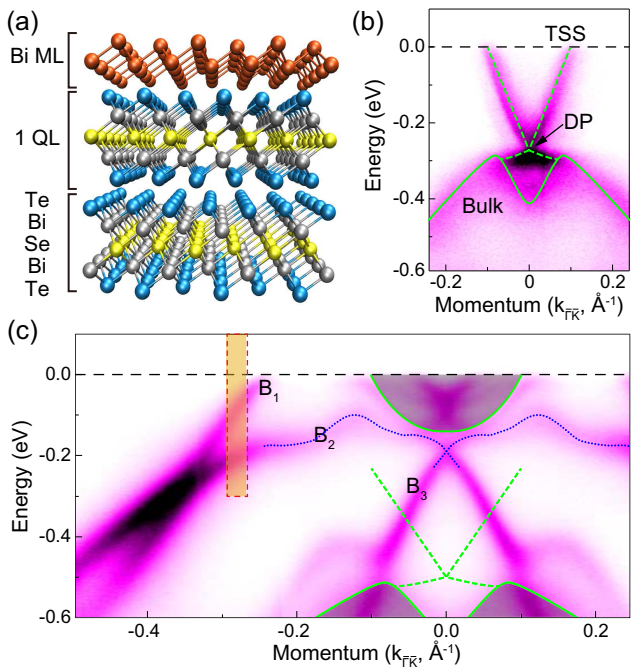}
\caption{Experimental band dispersions along the high symmetry direction $\bar{\Gamma}$--$\bar{K}$. (a) Atomic structure of a Bi monolayer on Bi$_2$Te$_2$Se. (b) Bulk (guided by solid line) and surface state (dashed lines) band dispersions of pristine Bi$_2$Te$_2$Se. (c) Band dispersions of Bi-covered Bi$_{2}$Te$_{2}$Se along the $\bar{\Gamma}$--$\bar{K}$ direction. The bulk bands of Bi$_2$Te$_2$Se are guided by solid lines and hatchings. The surface state of pristine Bi$_2$Te$_2$Se (dashed lines) disappears. The new surface states of B$_1$, B$_2$ and B$_3$ originate from the Bi monolayer.}
\end{figure}

Figure 1(b) is the band dispersion of a well-known 3D TI, BTS, whose atomic structure is depicted in Fig.~1(a). Its TSS shows the V-shape dispersion above the bulk valence band (dashed lines), exposing the characteristics of Dirac electrons. The Dirac point is located at $\sim$0.3~eV below the Fermi level as reported before \cite{NCom.3.636}. On top of fresh cleaved surfaces, we epitaxially grow atomically flat Bi(111) films \cite{PRB.89.155436}. The atomic structure of the monolayer film is shown in Fig.~1(a), which was confirmed by STM and \textit{ab initio} calculations \cite{PRB.89.155436}. The electronic band dispersions of BTS are substantially changed after growing a monolayer Bi film as shown in Fig.~1(c). Firstly, the bulk bands shift down by $\sim$0.2~eV due to the charge transfer from the Bi film as discussed further below. This makes the edge of the bulk conduction band appears below the Fermi energy around the $\bar{\Gamma}$~point. This behavior was previously observed for various electron-doping adsorbates on Bi$_2$Te$_3$ and related surfaces \cite{N.460.1101, PRL.104.016401, S.325.178, S.332.560, NCom.3.636}. Secondly, but most surprisingly, the TSS of BTS (green dashed lines) disappears completely and instead a $\Lambda$-shape band (B$_3$) emerges around $\bar{\Gamma}$. In addition, two strongly dispersing states appear away from $\bar{\Gamma}$; one (B$_1$) crosses the Fermi level to become metallic and the other (B$_2$) is connected to the $\Lambda$-shape band (B$_3$) at $\bar{\Gamma}$ forming a new band crossing (blue dashed lines). These data reveal greater details of the band dispersions with the improved spectroscopic resolution, which are overall consistent with the recent studies on a similar system of Bi/Bi$_2$Te$_3$ \cite{PRL.107.166801, PRL.109.016801}.

The experimental band dispersions match well with the \textit{ab initio} calculation, which further unveils the origin of the newly formed bands. The two largely dispersing bands (B$_1$ and B$_2$) are the $p_{xy}$ bands of the Bi monolayer. These bands are degenerate in the freestanding Bi(111) monolayer [Fig.~2(a)] but split due to the inversion symmetry breaking by and the interaction with the substrate. We can artificially control the strength of such an interaction in the calculation by changing the distance between the Bi film and the TI substrate. Even at a moderate interaction [Fig.~2(b)], the TSS and the 2D electronic states of the film are strongly hybridized to make B$_2$ and B$_3$ form a Rashba-type pair and B$_1$ form a Dirac cone above $E_F$ and disperse into the conduction band of BTS. A stronger hybridization [Fig.~2(c)] causes the TSS totally vanishes leaving only three bands within the BTS band gap (blue shades), B$_1$ and the B$_2$/B$_3$ Rashba pair. The Bi film becomes metallic due to the B$_1$ band dispersing into the BTS conduction band with its electrons partly transferred to the substrate. This explains the downward shift of the band of BTS. Surprisingly and unexpectedly, the Bi/BTS complex has three surface states of only the Bi origin within the BTS band gap with the original TSS removed.

\begin{figure}[!ptb]
\includegraphics{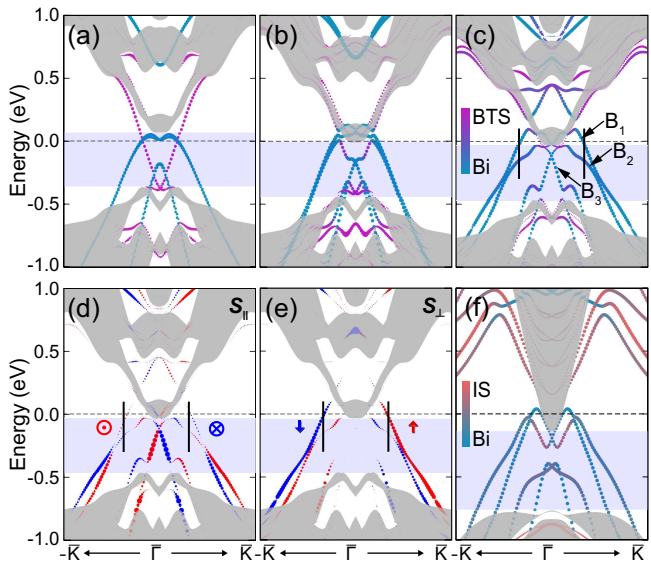}
\caption{Spin-resolved band structure calculations and their spin textures along $\bar{\Gamma}$--$\bar{K}$. Calculated band structure for a Bi monolayer on Bi$_2$Te$_2$Se along the $\bar{\Gamma}$--$\bar{K}$ direction (a) without and [(b) and (c)] with the film-substrate interaction. In (b) the Bi monolayer is detached from the substrate further from the equilibrium position of (c) by 2\AA. Gray colored region indicates the Bi$_2$Te$_2$Se bulk state while the band originating from the Bi film (Bi$_2$Te$_2$Se) is colored in cyan (magenta). [(d) and (e)] Spin orientations of the surface state bands shown in (c). Opposite in- and out-of-plane spin components are indicated by red and blue colored dots. The size of dots represent the magnitude of the corresponding spin components. (f) Calculated band structure of the Bi monolayer on a trivial insulator In$_2$Se$_3$. The blue shades indicate the band gap of the substrates.} 
\end{figure}

The strong spin-orbit coupling of Bi itself and the Rashba-type band crossing for the newly formed surface states (B$_2+$B$_3$) suggest that they are spin polarized. This spin polarization is detailed in the band calculation (Fig.~2). In particular, along the $\bar{\Gamma}$--$\bar{K}$ direction of the momentum space, where the band dispersion shown in Fig.~1 was measured, not only B$_2+$B$_3$ but also the B$_1$ band are fully spin polarized with largely different spin orientations. While B$_2+$B$_3$ bands have strong in-plane spin components within the surface plane, being consistent with the Rashba spin splitting, B$_1$ has its spin in the surface normal. While there is the strong out-of-plane spin component, the helical spin texture, opposite spins at opposite momenta, is obvious for all three surface states [Figs.~2(d), 2(e), 3(c), and 3(e)]. The partially out-of-plane spin polarization (B$_2$ in the present case) is widely found for normal TSS of 3D TI materials, for the momentum space away from the Dirac point, and is due to the warped band structure reflecting the hexagonal crystal structure \cite{PRL.103.266801}. However, in the present case, the B$_1$ spin at the Fermi energy is unusually dominated by the out-of-plane component, except for apexes of the hexagonal Fermi surface, where the spin flips [Fig.~3(c)]. The band dispersions and their detailed spin texture obviously show that the triple bands formed within the band gap are spin helical surface states between the BTS bulk and the vacuum. We thus conclude that they play the role of TSS connecting topologically nontrivial BTS with trivial vacuum, satisfying apparently \textit{the topological requirement of an odd number of spin helical surface states crossing the band gap}.

The spin polarization of the major bands was directly measured by SARPES. At the electron momenta away from $\bar{\Gamma}$, the two major surface state bands are prominent as shown in the energy distribution curves of photoelectrons [Figs.~3(a) and 3(b)], which were taken for the B$_1$ and B$_2$ bands at the momenta indicated in Figs.~1(c) (the dashed box) and 2 (vertical bars). The strong spin polarization is very clear in the SARPES data. As predicted in the calculation (Fig.~2), the spin is almost vertical, in particular for B$_1$, with opposite orientations between the opposite directions of electron momenta. The strong in-plane spin polarization of the B$_3$ band is also clearly confirmed (data not shown), which is fully consistent with a very recent report \cite{PRB.89.155116}. This result clearly evidences the spin helical nature of the three surface state bands. We measured a wide binding energy range and the agreement between the experiment and the calculation is excellent for the detailed spin orientation.

\begin{figure}[!pb]
\includegraphics{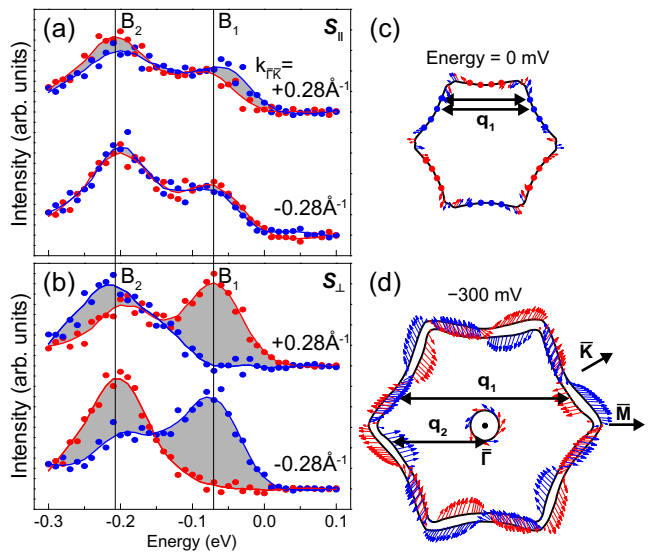}
\caption{Spin-polarized photoelectron intensity for the B$_1$ and B$_2$ bands for the (a) in- and (b) out-of-plane spin components measured by SARPES. The in-plane spin component measured is perpendicular to electron momenta, which is important for the spin helical system. The photoelectron energy distribution curves were taken at electron memento specified in Fig.~1(c) (the dashed box) and Figs.~2(c), 2(d) and 2(e) (vertical bars). Constant-energy contours and the spin texture of the surface state bands calculated at (c) the Fermi energy and (d) $-$300 meV. The red (blue) arrows and dots denote positive (negative) spin direction and the dots are for the dominating out-of-plane spin components. The possible back scattering wave vectors connecting parallel spins, \textit{$\textbf{q$_{1}$}$} and \textit{$\textbf{q$_{2}$}$}, are indicated.}
\end{figure}

The helical spin texture of the surface state is a hallmark of a TSS and it provides the unique and important property of suppressing electron backscattering. The electron backscattering can be directly confirmed by measuring the quasiparticle interference (QPI) in STM due to electron scatterings by defects. We performed STM experiments for partially covered Bi films composed of several 100~nm wide 2D islands~\cite{PRB.89.155436}. For these islands, we can observe the QPI pattern due the scattering by the edge of islands. The STS $dI/dV$ data in Fig.~4(a) shows the spatially-resolved local density of states (LDOS) containing such electron interference patterns (indicated by arrows). This QPI depends systematically on the electron energy reflecting the band structure. The Fourier transform of the QPI patterns reveals the scattering wave vectors involved \cite{N.466.343, PRL.103.266803}. Mainly two distinct scattering vectors are identified, \textit{$\textbf{q$_{1}$}$} and \textit{$\textbf{q$_{2}$}$}. They matches with those expected from the band structure and the spin texture discussed above; the electron backscattering is allowed only for the wave vectors connecting the spin parallel parts of the bands as shown in Figs.~3(c) and 3(d). This result, while not fully quantitative, further corroborates the calculation and the SARPES experiment. Consistent results were also obtained for the 2D QPI patterns taken at selected energies. 

\begin{figure}[!pt]
\includegraphics{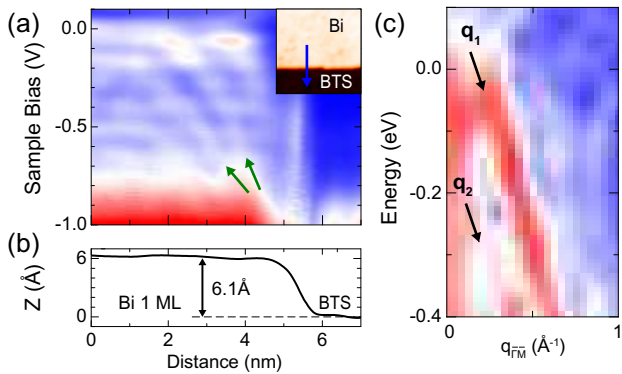}
\caption{Scatterings of surface state electrons measured by STM. (a) Spatially and energetically resolved STS ($dI/dV$) measurement taken across a step edge of a Bi monolayer island (along the blue arrow in the inset). The crystallographic orientation and the atomic structure of the step edges were clarified in the previous work \cite{PRB.89.155436}, which corresponds to the $\bar{\Gamma}$--$\bar{M}$ direction in momoentum space. The green arrows indicate the QPI pattern due to the electron scattering interference. (b) The topographic profile along the edge. (c) Energy-resolved Fourier transform of (a) shows electron scattering wave vectors, \textit{$\textbf{q$_{1}$}$} and \textit{$\textbf{q$_{2}$}$}, dispersing according the band structure given in Fig.~1(c). }
\end{figure}

By covering a single atom-thick Bi film on a 3D TI BTS, we show theoretically and experimentally, that the TSS of TI is totally removed and new spin helical electron states are formed in the covering film. This demonstrates that the TSS of a 3D TI can be transferred into different atomic layer covering the 3D TI. During this transfer, the band dispersion and the spin texture are substantially modified. In the present case, unique vertical spins are created. The underlying mechanism of the TSS transformation is the strong electronic hybridization between the metal monolayer and the TI including the TSS (Fig.~2). As a proof of the concept and the non-trival nature of the newly formed surface states, we performed similar calculations for a Bi film on a trivial insulator, In$_2$Se$_3$ [Fig.~2(f)]. This calculation yields only trivial Rashba bands of the Bi origin, four surface states paired, within the band gap. The basic concept seems similar to the change of the vertical positions of TSS for the insulating compound films covering a TI in the recent theoretical studies \cite{acsnano.6.2345,tuning, band_engineering}. Those theoretical suggestions were not realized yet and our choice of simple and single-atom-thick film made the realization and confirmation of the TSS transformation substantially easier.

The formation mechanism is shown to be general since our calculation for the Sb monolayer leads to the same situation. However, a Ge film does not work to replace the TSS since its bands do not overlap with the TSS preventing any electronic hybridization within the band gap. We also tested various different TI substrates, which all yield consistent results. The only difference between different substrates is the degree of hybridization of Bi and the substrates, especially for the band near the Rashba crossing. When the Rashba crossing gets closer to the conduction band minimum of the substrate, the corresponding wave functions absorb more substrate character. This different degree of the hybridization is consistent with the recent report of 'hybridized Dirac cones' for Bi/Bi$_2$Se$_3$ and Bi/Bi$_2$Te$_3$, respectively \cite{PRB.89.155116}. Note that this crossing is not a Dirac state but a Rashba crossing and the trivial case of Bi/In2Se3 also has such hybridized Rashba crossing near the conduction band minim [Fig. 2(f)]. What is quintessential in the topological band texture of these systems is, thus, not the presence of the Rashba band crossing, but the odd number of spin helical bands within the band gap. The hybridization of the B1 band with the conduction bands of the substrate makes this  \textit{odd-numbered} band texture [Fig.~2(b)].

In summary, we show that the TSS of a TI can be completely replaced by or transformed into two dimensional spin helical electronic states of a strongly interacting monolayer covering the surface, through a strong electronic hybridization. In principle, with properly designed covering films, one can make various different types of TSS on a single 3D TI, providing a new degree of freedom in creating topologically protected electron channels beyond the limit of the bulk materials synthesis, with possibtly tailored properties for various TI applications.

This work was supported by Institute for Basic Science (IBS) through the Center for Artificial Low Dimensional Electronic Systems and SRC Center for Topological Matter (2011-0030046), and the Basic Science Research program (2012-013838).

\end{document}